\documentclass{article}

\usepackage{amsmath}
\usepackage{amssymb}
\usepackage{simplemargins}
\usepackage{graphicx}
\usepackage{color}
\usepackage{epstopdf}

\settopmargin{2cm}
\setleftmargin{2cm}
\setrightmargin{2cm}
\setbottommargin{2cm}

\newcommand{\review}[1] {#1}
\newcommand{\reviewtwo}[1] {#1}

\newenvironment{enumerate*}%
  {\begin{enumerate}%
    \setlength{\itemsep}{0pt}%
    \setlength{\parskip}{0pt}}%
  {\end{enumerate}}

\begin{document}
\title{Efficient parametric inference for stochastic biological systems with measured variability}
\author{Iain G. Johnston\\ \normalsize Department of Mathematics, Imperial College London, UK SW7 2AZ}
\date{}

\maketitle

\begin{abstract}
Stochastic systems in biology often exhibit substantial variability within and between cells. This variability, as well as having dramatic functional consequences, provides information about the underlying details of the system's behaviour. It is often desirable to infer properties of the parameters governing such systems given experimental observations of the mean and variance of observed quantities. In some circumstances, analytic forms for the likelihood of these observations allow very efficient inference: we present these forms and demonstrate their usage. When likelihood functions are unavailable or difficult to calculate, we show that an implementation of approximate Bayesian computation (ABC) is a powerful tool for parametric inference in these systems. However, the calculations required to apply ABC to these systems can also be computationally expensive, relying on repeated stochastic simulations. We propose an ABC approach that cheaply eliminates unimportant regions of parameter space, by addressing computationally simple mean behaviour before explicitly simulating the more computationally demanding variance behaviour. We show that this approach leads to a substantial increase in speed when applied to synthetic and experimental datasets.
\end{abstract}

\section{Introduction}
Random processes and variability in cellular biology have been the focus of much recent study, with increasing evidence that variability in biological quantities within and between cells influences life on a remarkable number of levels. Examples include within- and between-cell noise in gene expression \cite{elowitz2002stochastic, kaern2005stochasticity, blake2003noise, raser2004control, paulsson2005models, rausenberger2008quantifying}, variability between cells in organelle and energetic content \cite{johnston2012mitochondrial, neves2010connecting}, stem cell fate decisions \cite{chang2008transcriptome, enver2009stem, graf2008heterogeneity}, bacterial strategies \cite{kussell2005bacterial} and growth and drug response of cancer cells \cite{brock2009non, spencer2009non}. In addition to this physiological importance, variability in biological measurements represents an often unexplored source of information about underlying biological mechanisms, as magnitudes of variance measurements allow more powerful inference of the parameters governing the emergence of variability in stochastic biological systems. 

Deduction of quantitative descriptions of such systems falls within the field of parametric inference \cite{wilkinson2012stochastic}. In a typical inference problem, a model of the system is constructed and parameterised with some trial parameter values, possibly drawn from some prior probability distribution. If analytic solutions for the model's behaviour are unavailable, as is usually the case with all but the simplest biological systems, the model is then simulated and its behaviour is compared to data from biological experiments. However, simulation of stochastic models (usually through the Gillespie algorithm \cite{gillespie1977exact} or variants) is often computationally expensive, requiring many calculations to adequately characterise the model's behaviour. It is therefore desirable to minimise the number of stochastic simulations required to perform parametric inference for noisy biological systems.

In this study, we consider biological systems where the available experimental measurements are statistics of a population of individuals -- for example, the mean and variance of some property \reviewtwo{across a number of cells} -- rather than focussing on individual measurements \cite{wilkinson2012stochastic}. We first discuss situations in which analytic expressions for the expected values and variances of these statistics allows very fast parametric inference. Cases in which analytic solutions are unavailable are then considered, using approximate Bayesian computation (ABC), a means of performing parametric inference without computing explicit likelihoods \cite{toni2009approximate, beaumont2002approximate, marin2012approximate, sunnaker2013approximate}. A modification to this protocol is introduced, dramatically speeding up such inferential tasks by decreasing the number of required stochastic simulations. We illustrate the use of all these approaches with a theoretical case study, and the use of the proposed ABC implementation by analysing experimental data on gene expression after transcription induction.

\section{Analytic likelihoods associated with measurements of \review{mean and variance}}
In the work below, we will assume that experimental measurements exist of the mean and variance of some quantity of biological interest. The existence of an appropriate stochastic model to describe the system is also assumed.

We will begin by considering the likelihood associated with a particular model parameterisation given measurements of $m$, the mean of $n$ experimental realisations all measured at the same time (or other co-ordinate) $t$, and of the variance $v$ of these $n$ quantities. For example, we may imagine having a \review{system} of $n$ cells, and at some time $t$ measurements of RNA transcript number are taken in each of these cells. If we label each individual cell measurement $x_i$, where $i$ labels the corresponding cell, then the mean transcript number over these $n$ measurements would then constitute our measured \review{sample} mean $m = n^{-1} \sum_i x_i$, and the variance over the $n$ measurements would constitute the \review{sample} variance $v = (n-1)^{-1} \sum_i (x_i - m)^2$. In general, measurements at many time points $t$ may be taken, so that we develop time series for \review{sample} statistics $m$ and $v$.

An underlying model for such a system will be represented with the form $x(t, \theta) = M(t, \theta) + \eta(\theta)$, so that each individual quantity $x$ depends on a model function $M$ of an ordinate $t$ and parameterisation $\theta$, and includes additive noise $\eta(\theta)$. We will assume that noise is uncorrelated and has zero mean.

If several conditions are met, analytic expressions may be obtained for the likelihoods associated with a dataset of sampled $m$ and $v$ statistics. In the below, we assume that the $n$ samples contributing to each recorded statistic are drawn from an underlying Normal distribution with known variance $\sigma^2$: that is, $\eta$ is uncorrelated and Normally distributed. We also require that this number of samples $n$ is known. Then (and generally), the expected value of the \review{sample} mean $m$, under our model, is $\mathbb{E}(m) = \mu = M(t, \mathbf{\theta})$, the deterministic model term, as the expected sum of samples from the noise term is zero. Since, in this case, the distribution of the sampled mean can be assumed to be Normal, the variance of this estimate is then $\mathbb{V}(m) = \sigma^2 / n$. Overall, the log-likelihood associated with an observed \review{sample} mean $m$ is:

\begin{equation}
\mathcal{L}(\mathbf{\theta} | m) = \frac{-1}{2} \log \left( 2 \pi \sigma^2 / n \right) - \frac{(m - \mu)^2}{2 \sigma^2 / n}. \label{likmean}
\end{equation}

As we assume that the $n$ individual measurements are normally distributed, the sample variance follows a scaled $\chi^2$ distribution (see, for example, Proposition 2.11 in Ref. \cite{knight2000mathematical}): specifically, \review{$v (n-1) / \sigma^2 \sim \chi^2_{n-1}$}. The expected sample variance is thus $\mathbb{E}(v) = \sigma^2$, and the variance of the sample variance is $\mathbb{V}(v) = \frac{2 \sigma^4}{n-1}$. From the underlying $\chi^2$ distribution, the log-likelihood associated with a variance measurement $v$ is:

\begin{equation}
\mathcal{L}(\mathbf{\theta} | v) = \frac{1}{2} \left( (n-3) \log \left( \frac{v (n-1)}{\sigma^2} \right) - (n - 1) \left( \frac{v}{\sigma^2} + \log 2 \right) - 2 \log \Gamma \left( \frac{n-1}{2} \right) \right) \label{liknormal}
\end{equation}


The sum of the log-likelihoods in Eqns. \ref{likmean} and \ref{liknormal} then gives the overall log-likelihood associated with the observed $m$ and $v$ statistics. In general, $\mu$ and $\sigma^2$ are functions of time: the log-likelihood associated with a time series of recorded $m$ and $v$ statistics is then the sum of Eqn. \ref{likmean} and \ref{liknormal} computed at each timepoint for which measurements exist.

\section{ABC, MCMC, and comparing data to simulation}
In cases where details of the measurement protocol are absent, or the measurements cannot be assumed to come from an underlying Normal distribution, the likelihood function associated with measurements of \review{sample} statistics is likely to be intractable. In these cases, we propose employment of the protocol of approximate Bayesian computation (ABC) \cite{toni2009approximate, beaumont2002approximate, marin2012approximate, sunnaker2013approximate}. ABC avoids explicit computation of likelihoods by employing a simpler comparison between data and simulation, the posteriors from which converge, in the limit of strict comparison between data and simulation, on the true posteriors. It should be noted, however, that the absence of explicit likelihoods does not come without costs in accuracy \cite{sunnaker2013approximate}.

In basic implementations, the comparison measure used in ABC takes the form of a distance measure between simulated data $\mathcal{D'}(\mathbf{\theta})$ and observed data $\mathcal{D}$ (for example, the Euclidean distance between the datasets), where we have explicitly written the simulated data as a function of a trial parameterisation $\mathbf{\theta}$. In situations with larger or complicated datasets, statistics are used to summarise the data, and the distance between simulated and observed summary statistics $S(\mathcal{D'}(\mathbf{\theta}))$ and $S(\mathcal{D})$ are used. If the distance between the data (or summary) from observation and that from simulation is zero, or below a certain threshold $\epsilon$, the parameters $\mathbf{\theta}$ from that simulation are recorded as a sample from the posterior. In this way we build up a posterior distribution on $\mathbf{\theta}$ labelled by the threshold $\epsilon$ employed: $P_{ABC}(\mathbf{\theta} | \rho(\mathcal{D'}(\mathbf{\theta}), \mathcal{D}) \leq \epsilon)$. As the acceptance threshold $\epsilon$ is decreased (forcing a stricter agreement with the data), this computed posterior converges on the true posterior:

\begin{equation}
P_{ABC}(\mathbf{\theta} | \rho(\mathcal{D'}(\mathbf{\theta}), \mathcal{D}) \leq \epsilon) \xrightarrow{\epsilon \rightarrow 0} P(\mathbf{\theta} | \mathcal{D}).
\end{equation}

We are left with the problem of choosing a suitable distance metric with which to compare our model to our observed data. Given that $\mathcal{D} = \{ t_i, m_i, v_i \}$ are the observed data and $\mathcal{D'}(\mathbf{\theta}) = \{ \mu(t | \mathbf{\theta} ) , \sigma^2(t | \mathbf{\theta} ) \}$ are the simulated mean and variance trajectories, we propose the following distance function:

\reviewtwo{
\begin{equation}
\rho(\mathcal{D'}(\mathbf{\theta}), \mathcal{D}) = \sum_{\text{datapoints}\,i} (\log \mu(t_i | \mathbf{\theta}) - \log m_i)^2 + (\log \sigma^2(t_i | \mathbf{\theta}) - \log v_i)^2, \label{distancemeasure}
\end{equation}
}

where $\mu(t_i | \mathbf{\theta})$, $\sigma^2(t_i | \mathbf{\theta})$ are respectively the mean and variance measurements at time $t_i$ from the simulated dataset $\mathcal{D'}(\mathbf{\theta})$. \reviewtwo{This simulated dataset should be obtained by performing $n$ repeats of the required stochastic simulation, where $n$ is the number of individual elements that gave rise to the corresponding experimental measurement (for example, individual cells).} Overall, we therefore have the sum of squared differences between the predicted and observed mean and variance, taken in logarithmic space to facilitate a comparison of the multiplicative rather than additive differences (and thus allow a comparison across the potentially different magnitudes of means and variances). If a particular measurement has an associated mean but not a variance measurement (or, pathologically, vice versa), the missing value is ignored. Minimising this distance is then equivalent to performing least-squares regression in logarithmic space for the trajectories of mean and variance behaviour of our system.

This distance measure thus includes all available information about the means and variances observed in the data and compares them multiplicatively to the means and variances predicted in the model, comparing mean and variance data with equal weighting. We note that in situations where the contribution from the set of either mean or variance measurements is known to always produce a higher distance than measurements from the other set, this weighting could be changed to allow more efficient exploration of potential model parameterisations (see Discussion).

We now consider how to implement this approach to perform Bayesian inference. We begin with a prior distribution $\pi(\mathbf{\theta})$ over parameters $\mathbf{\theta}$ and wish to obtain a posterior distribution over these parameters. A basic ABC rejection algorithm simply samples trial parameterisations from the prior distribution, accepting these as posterior samples if their distance (computed via stochastic simulation) from the observed data falls below $\epsilon$. A framework that has been shown to be more efficient involves embedding the ABC process in a Markov chain Monte Carlo (MCMC) framework \cite{marjoram2003markov}. Here, rather than randomly sampling trial parameterisations from the prior distribution, a small perturbation is made to the previous trial parameterisation to yield a new trial, ensuring that regions of parameter space which have previously yielded accepted parameterisations are explored preferentially. There exists a great range of possible perturbation protocols: we will represent a perturbation kernel as $q( \mathbf{\theta} \rightarrow \mathbf{\theta'} )$, the probability of proposing new parameterisation $\mathbf{\theta'}$ given existing parameterisation $\mathbf{\theta}$. In the work below, we will generally employ a Gaussian perturbation kernel on model parameters, with variance chosen to yield an acceptance rate of around $50\%$ (see below).

This MCMC approach requires an initialisation step -- that is, identifying the first parameterisation in the chain -- which can be implemented using random sampling from the prior or an optimisation approach to find a suitable start point. We can use this framework and our distance measure in an ABC algorithm as detailed below.

\vspace{0.3cm}
\textbf{Algorithm 1 -- ABC MCMC for mean and variance measurements}
\begin{enumerate*}
\item Choose a rejection threshold $\epsilon$ 
\item Do \emph{(initialisation)}
\begin{enumerate*}
\item Pick \emph{(depending on desired search method)} a new random parameterisation $\mathbf{\theta'}$ from prior
\item Perform stochastic simulation \reviewtwo{$n$ times} to compute $\rho = \rho(\mathcal{D'}(\mathbf{\theta'}), \mathcal{D})$ \emph{(Eqn. \ref{distancemeasure})}
\end{enumerate*}
\item while $\rho > \epsilon$ 
\item $\mathbf{\theta} = \mathbf{\theta'}$
\item Do \emph{(sampling)}
\begin{enumerate*}
\item Apply a perturbation to $\mathbf{\theta}$ according to a transition kernel $q( \mathbf{\theta'} \rightarrow \mathbf{\theta} )$ to obtain a new parameterisation $\mathbf{\theta'}$
\item Perform stochastic simulation \reviewtwo{$n$ times} to compute $\rho = \rho(\mathcal{D'}(\mathbf{\theta'}), \mathcal{D})$
\item If $\rho \leq \epsilon$ then set $\theta = \theta'$ with probability $\min \left( 1, \frac{\pi (\mathbf{\theta'}) q( \mathbf{\theta'} \rightarrow \mathbf{\theta}) }{\pi (\mathbf{\theta}) q( \mathbf{\theta} \rightarrow \mathbf{\theta'} ) } \right) $
\item Record $\theta$
\end{enumerate*}
\item while(termination condition not met)
\end{enumerate*}

In the initialisation step, new parameterisations may be picked simply according to the prior distribution, or using simulated annealing or other heuristic search methods to increase the efficiency of the search for a suitable initial parameterisation. The sampling part of Algorithm 1 follows the form of Algorithm F with alteration F3' from Ref. \cite{marjoram2003markov}. We note that if uniform priors are used and the transition kernel $q$ is symmetric, the ratio in the acceptance probability for parameterisations with $\rho \leq \epsilon$ is simply unity.

\section{Mean-first ABC: an efficient two-step ABC algorithm}
In applications of MCMC in parametric inference, a heuristic approach to ensure a combination of an adequate search of parameter space and a reasonable number of samples from a posterior distribution is that approximately 50\% of proposed parameterisations should be accepted. If this heuristic is employed for a complicated stochastic system, then, without algorithmic refinements, half of the stochastic simulations performed will not contribute to final samples from the posterior distribution. It is thus desirable to find a fast way of determining whether a given proposed parameterisation is destined to be rejected.

In stochastic biological systems, it is often the case that the mean behaviour of a system is calculable without explicit stochastic simulation. The mean behaviour in these cases may be obtained analytically for simple systems or by numerically solving a set of ODEs for more complicated systems: both these approaches are generally computationally easier than performing an ensemble of stochastic simulations. Here we propose an algorithm that takes advantage of this easy calculability to avoid spending computational resources performing stochastic simulations in regions of parameter space that do not produce adequately comparable mean behaviour, thus dramatically decreasing the required resources required to sample posterior distributions.

Assume we have mean data $\mathcal{M} = \{(t_i, m_i)\}$ and variance data $\mathcal{V} = \{(t_i, v_i)\}$. In the algorithm below, two types of computation are performed. The first employs \review{an analytic function $\hat{\mu}(t | \mathbf{\theta})$ for the expected value of a quantity} at time $t$ given parameters $\theta$. \review{We calculate}

\review{
\begin{equation}
\hat{\rho}_m(\mathcal{D'}(\mathbf{\theta}), \mathcal{D}) = \sum_{\mbox{\tiny datapoints}\, i} (\log(\hat{\mu}(t_i | \mathbf{\theta})) - \log(m_i))^2 \label{detrhomean}.
\end{equation}
}

The second employs repeated stochastic simulations to characterise \review{the mean $\mu$ and variance $\sigma^2$ of a quantity} at time $t$. We then construct

\review{
\begin{eqnarray}
\rho_m(\mathcal{D'}(\mathbf{\theta}), \mathcal{D}) & = & \sum_{\mbox{\tiny datapoints}\, i} (\log(\mu(t_i | \mathbf{\theta})) - \log(m_i))^2 \label{stochrhomean} \\
\rho_v(\mathcal{D'}(\mathbf{\theta}), \mathcal{D}) & = & \sum_{\mbox{\tiny datapoints}\, i} (\log(\sigma^2(t_i | \mathbf{\theta})) - \log(v_i))^2 \label{stochrhovariance}.
\end{eqnarray} 
}







 
\review{The computation of $\hat{\rho}_m$ is cheap; the computation of $\rho_m$ and $\rho_v$ is expensive}, \reviewtwo{relying on repeated stochastic simulation}. \review{We can then take advantage of the fact that if $\hat{\rho}_m > \epsilon$, it is highly likely that $\rho_m + \rho_v > \epsilon$. In other words, if the discrepancy between the deterministic mean trajectory and experimental measurements exceeds a threshold, it is likely that the combination of discrepancies from the sample mean trajectory and the sample variance trajectory will exceed this threshold. This intuitive result can be shown to apply if mean measurements involve reasonable numbers of datapoints (so that the deterministic and sample means behave similarly) and if discrepancies related to mean trajectories are broadly similar or lower than those related to variance trajectories (see Appendix).}

\review{To implement this idea, we first compute $\hat{\rho}_m$ using Eqn. \ref{detrhomean}. If $\hat{\rho}_m > \epsilon$, we already know that this trial parameterisation likely falls outside the ABC acceptance threshold, and can immediately reject this step. Only if $\hat{\rho}_m < \epsilon$ do we perform the expensive stochastic simulation to determine $\rho_m$ and $\rho_v$ and thus check whether $\rho_m + \rho_v < \epsilon$.} In this manner, we use a bound that is computationally cheap \review{to calculate to} exclude inappropriate regions of parameter space, in a process conceptually similar to techniques used in optimisation of data mining protocols \cite{ding2008querying}. The algorithm implementing this approach follows below.

\vspace{0.3cm}

\review{
\textbf{Algorithm 2 -- Mean-first ABC with MCMC}
\begin{enumerate*}
\item Choose a rejection threshold $\epsilon$. 
\item Do \emph{(initialisation)}
\begin{enumerate*}
\item Pick \emph{(depending on desired search method)} a new random parameterisation $\mathbf{\theta'}$
\item Compute $\hat{\rho}_m = \hat{\rho}_m(\mathcal{D'}(\mathbf{\theta'}), \mathcal{D})$
\item If $\hat{\rho}_m \leq \epsilon$, perform stochastic simulation \reviewtwo{$n$ times} to obtain $\rho_m = \rho_m(\mathcal{D'}(\mathbf{\theta'}), \mathcal{D})$, $\rho_v = \rho_v(\mathcal{D'}(\mathbf{\theta'}), \mathcal{D})$, else $\rho_m, \rho_v = \infty$
\end{enumerate*}
\item while $\hat{\rho}_m > \epsilon$ and $\rho_m + \rho_v > \epsilon$
\item $\mathbf{\theta} = \mathbf{\theta'}$
\item Do \emph{(sampling)}
\begin{enumerate*}
\item Apply a perturbation to $\mathbf{\theta}$ to obtain a new parameterisation $\mathbf{\theta'}$
\item Compute $\hat{\rho}_m = \hat{\rho}_m(\mathcal{D'}(\mathbf{\theta'}), \mathcal{D})$
\item If $\hat{\rho}_m \leq \epsilon$, perform stochastic simulation \reviewtwo{$n$ times} to obtain $\rho_m = \rho_m(\mathcal{D'}(\mathbf{\theta'}), \mathcal{D})$, $\rho_v = \rho_v(\mathcal{D'}(\mathbf{\theta'}), \mathcal{D})$, else $\rho_m, \rho_v = \infty$
\item If $\rho_m + \rho_v < \epsilon$ then set $\theta = \theta'$ with probability $\min \left( 1, \frac{\pi (\mathbf{\theta'}) q( \mathbf{\theta'} \rightarrow \mathbf{\theta} )}{\pi (\mathbf{\theta}) q( \mathbf{\theta} \rightarrow \mathbf{\theta'} )} \right)$
\item Record $\theta$
\end{enumerate*}
\item while(termination condition not met)
\end{enumerate*}
}

\section{Theoretical case study: a birth-death process}
We first compare the techniques described above with a theoretical case study. Here, we consider a birth-death process: that is, a system containing a number of elements $x$ which may replicate ($x \rightarrow 2x$) and degrade ($x \rightarrow 0$) according to Poisson processes with rates $\lambda$ and $\nu$ respectively. This abstract model may be used, for example, to describe copy numbers of bacterial population, or subcellular entities (e.g. mitochondria or chloroplasts) which divide and degrade randomly.

\begin{figure}
\includegraphics[width=9cm]{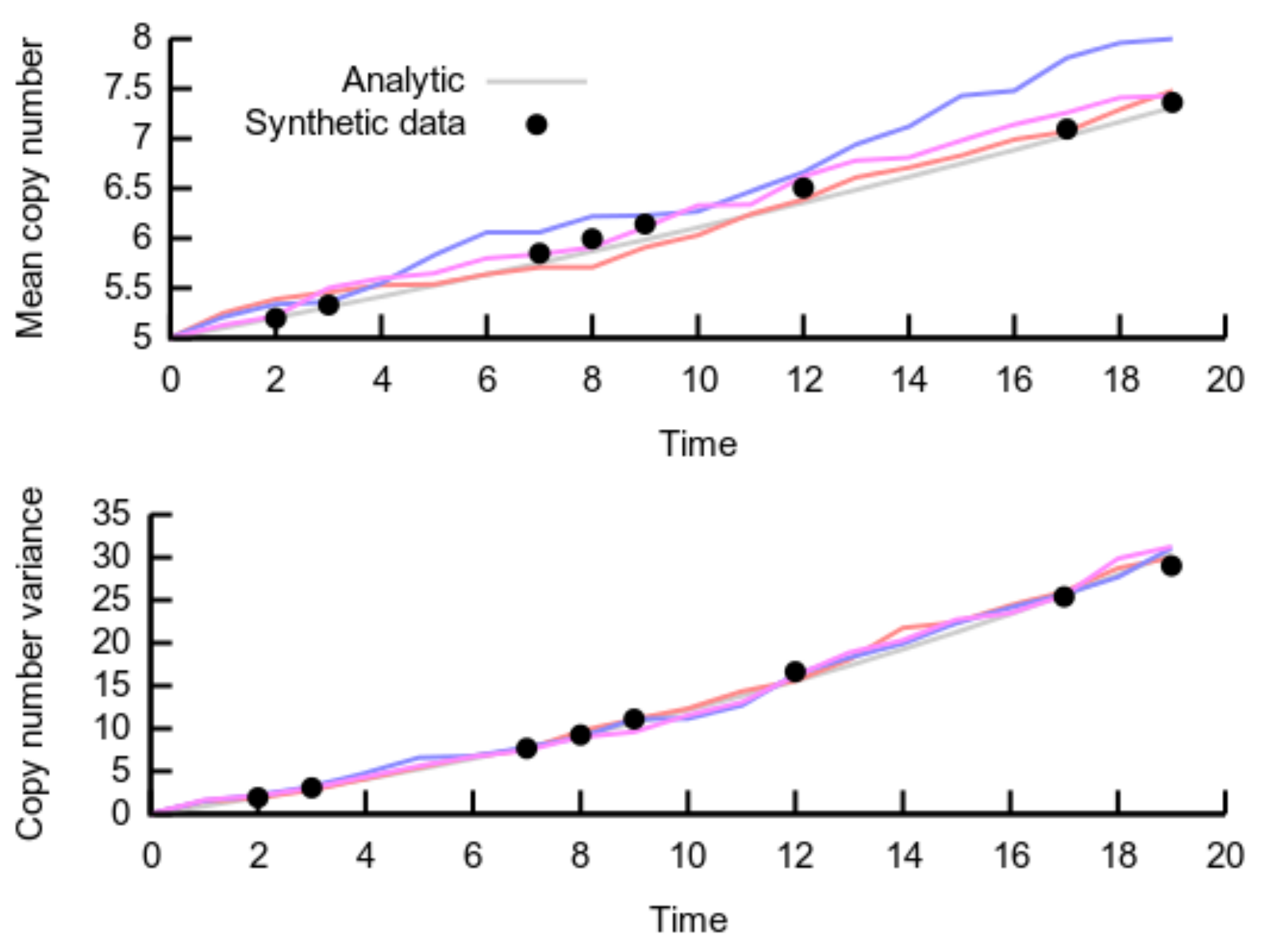}
\reviewtwo{\caption{\textbf{Synthetic dataset and analytic expressions for the birth-death model.} (top) Mean and (bottom) variance of a birth-death model with the parameterisation shown in the main text. Points show synthetic measurements used in the illustrative case study; grey lines show analytically-derived trajectories for the birth-death model; coloured lines show example trajectories with $\epsilon < 0.1$ resulting from parameterisations in the ABC inference simulations.}}
\label{figsynthtraj}
\end{figure}

To illustrate the application of the above approaches for parametric inference, we synthesise test data from the birth-death process with $\lambda = 0.1$, $\nu = 0.08$, $x_0 = 5$. We use the Gillespie algorithm \cite{gillespie1977exact} to realise \reviewtwo{$n = 100$} stochastic trajectories of the model and record measurements of \review{sample} statistics at a randomly-chosen set of times. These measurements (illustrated in Fig. \ref{figsynthtraj}) form the synthetic dataset used in this section. 

We begin by noting that without measurements of the variance in this system, the absolute magnitudes of $\lambda$ and $\nu$ cannot be determined: it is only their difference $\lambda - \nu$ that governs the mean behaviour of the process (see Eqn. \ref{eqnbdmean}). Considering observations of the variance, which also depends on the absolute values of these parameters (see Eqn. \ref{eqnbdvar}), parametric inference is more powerful.

\reviewtwo{\textbf{Analytic approximation for likelihood.}} With this simple birth-death model, we can obtain exact results for all moments of the copy number distribution $\mathbb{P}(x,t)$ using a generating function analysis \cite{bailey1964elements}. We proceed quickly through this process by noting that the probability generating function $G(z,t) = \sum_x z^x P(x,t)$ for the birth-death process is well known (see, for example, Section 8.6 of Ref. \cite{bailey1964elements}):

\begin{equation}
G(z,t) = \left( \frac{(z-1) \nu e^{(\lambda-\nu)t} - \lambda z + \nu}{(z-1) \lambda e^{(\lambda-\nu)t} - \lambda z + \nu} \right)^{x_0},
\end{equation}
and that we can use the standard relations $\mathbb{E}(x,t) = \left. \partial G / \partial z \right|_{z=1}$ and $\mathbb{V}(x,t) = \left. \partial^2 G / \partial z^2 \right|_{z=1} + \mathbb{E}(x,t) - \mathbb{E}(x,t)^2$ to derive the following expressions for the moments of $\mathbb{P}(x,t)$:

\begin{eqnarray}
\mu = \, \mathbb{E}(x,t) & = & x_0 e^{(\lambda-\nu)t} ; \label{eqnbdmean}\\
\sigma^2 = \, \mathbb{V}(x,t) & = & x_0 (e^{\lambda t} - e^{\nu t}) e^{(\lambda-2\nu)t} \left( \frac{\lambda+\nu}{\lambda-\nu} \right).  \label{eqnbdvar}
\end{eqnarray}

We then insert the values of $\mu$ and $\sigma^2$ from Eqns. \ref{eqnbdmean} and \ref{eqnbdvar} into Eqns. \ref{likmean} and \ref{liknormal} to compute the exact likelihoods $\mathcal{L}$ associated with each measurement of the \review{mean and variance}, assuming that the copy number distribution is Normal. This assumption is reasonable for copy number distributions with low probability at the limiting value of zero: however, as the spread of copy numbers increases with time in this model, the assumption may become less valid for later measurements\reviewtwo{, so the posteriors derived from this treatment should be regarded as an approximation to the true posteriors.} 

\begin{figure*}
\includegraphics[width=18cm]{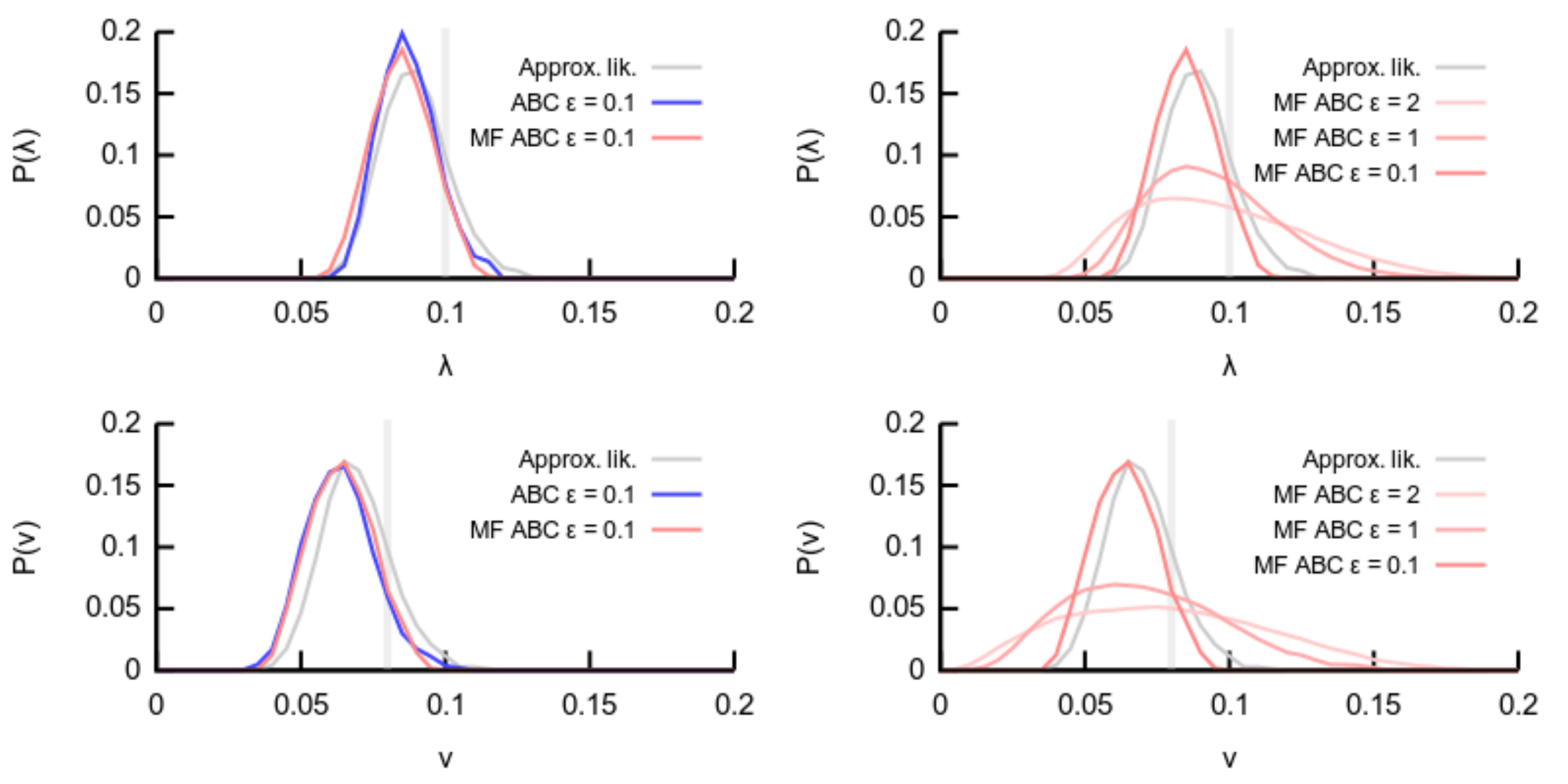}
\caption{\reviewtwo{\textbf{{Results for inference with synthetic data from approaches involving analytic approximation to the likelihood, ABC, and mean-first ABC.}} (left) Posteriors on (top) $\lambda$ and (bottom) $\nu$, the rate parameters of a birth-death model, using an analytic expression for the likelihood (assuming a Normal distribution of copy number; see text); ABC; and mean-first ABC. Posteriors from all approaches share similar structure. (right) Posteriors on (top) $\lambda$ and (bottom) $\nu$ as the mean-first ABC threshold $\epsilon$ is decreased. Thick vertical lines show the true parameter values underlying the simulations from which the data was produced.}}
\label{figmcmc}
\end{figure*}

To facilitate a comparison with the ABC approach, we use MCMC to derive posteriors using these analytic likelihood expressions (see below). The \review{posteriors derived with this approach} give strong support to the true underlying parameters (see Fig. \ref{figmcmc} and Table \ref{simstable}).


\textbf{ABC and mean-first ABC.} As an alternative to computing explicit likelihoods, we now employ ABC with the summed-log-squares distance measure in Eqn. \ref{distancemeasure}, using Algorithm 1 above. \reviewtwo{We now do not have to calculate explicit expressions for likelihood values.} We also employ this paper's refinement to the ABC algorithm by using Algorithm 2. 

Posteriors derived using these ABC approaches are presented in Fig. \ref{figmcmc}. Firstly, we see that the mean-first ABC yields the same posterior structure (within sampling noise) as the bare ABC posterior, as expected, as the mean-first algorithm only rejects parameterisations guaranteed to be rejected by the bare algorithm (though this implementation could be modified; see Discussion). \reviewtwo{We also illustrate the effects of varying the acceptance threshold $\epsilon$: reducing $\epsilon$ leads to convergence of the mean-first ABC-derived posterior to a form that both matches the ABC posterior with low $\epsilon$ and resembles the posterior derived using the approximate analytic likelihood.}

\textbf{Computational implementation and speed comparison.} The perturbation kernel in all these simulations was chosen to enforce an acceptance rate of around $50\%$. The initial parameterisation in each case was $\lambda = 0.2, \nu = 0.19, x_0 = 10$, a substantial distance away from the true parameterisation, and \review{we use uniform priors} $\lambda \sim U(0,0.5)$, $\nu \sim U(0,0.5)$, $x_0 = U(0, 100)$ (continuous, continuous, and discrete, respectively). \review{These priors were chosen to include a range far exceeding the parameter values for which reasonable support was expected, making no value more likely than any other within this wide range, and enforcing a non-negativity restriction on the parameters.} Sampling in every case was performed \reviewtwo{using $5 \times 10^6$ MCMC steps (following the preliminary part of the algorithm for the ABC simulations), of which the first $10\%$ consituted a `burn-in' period and the latter yielded posterior samples.} Each approach investigated yielded $100\%$ posterior density on $x_0 = 5$, unsurprisingly as the synthetic data intuitively provide strong support for this parameter (see Fig. \ref{figsynthtraj}).

\reviewtwo{In this case study we find that the mean-first protocol allows us to reject around $40\%$ of proposals at the lowest $\epsilon$ value used} without performing any Gillespie simulations, as the straightforwardly computable mean trajectories of these proposed parameterisations exceeded the acceptance threshold. This proportion corresponds to the mean-first protocol being responsible for \reviewtwo{around $70\%$} of rejections. 

In Table \ref{simstable}, we compare the computational resources required using each of these approaches. The mean-first ABC protocol leads to a substantial speedup in both the preliminary initialisation and sampling phases of the inference procedure, and thus represents the fastest tested approach for parametric inference without using analytic results for the likelihoods associated with \review{sample} statistics. \reviewtwo{To compare results from the different approaches, we record symmetrised Kullback-Leibler divergences between posteriors arising from using the analytic likelihood approximation, using bare ABC, and using mean-first ABC in Table \ref{simstable}. We find no substantial differences in the posteriors derived from bare ABC and mean-first ABC, both of which agree with the approximate likelihood treatment, and conclude that mean-first ABC allows a substantial computational speedup without a pronounced loss in accuracy.}



\begin{table}
\reviewtwo{
\footnotesize
\begin{tabular}{p{3.5cm}|p{2cm}|p{2cm}|p{2cm}|p{2cm}|p{2cm}}
& Needs analytic expression? & Preliminary simulations & MF rejections during sampling & \review{KL divergences for $\lambda$} & \review{KL divergences for $\nu$} \\
\hline
Approximate likelihood & Yes & - & - & 0, 0.1 & 0, 0.2 \\
Bare ABC $\epsilon = 0.1$ & No & $4 \times 10^3$ & - & 0.1, 0 & 0.2, 0 \\
MF ABC $\epsilon = 0.1$ & No & $4 \times 10^3$  & $ 40 \%$ & 0.08, 0.08 & 0.1, 0.1 \\
MF ABC $\epsilon = 1$ & No & $2 \times 10^3$ & $ 15 \%$ & 0.4, 1.2 & 0.9, 1.1 \\
MF ABC $\epsilon = 2$ & No & $1 \times 10^2$ & $ 6 \%$ & 1.3, 2.4 & 1.9, 2.3 \\
\hline
\end{tabular}
\caption{\textbf{Comparison of inference approaches.} Comparison of the posteriors and simulation statistics from inference using a likelihood-based approach and differently-parameterised ABC approaches. \review{Preliminary simulations: number of preliminary Gillespie runs (each consisting of $n$ trajectories) required to identify a trial parameterisation yielding residual under $\epsilon$. MF rejections: the mean proportion of fast rejections due to the mean-first (MF) protocol in the sampling stage of the inference process. Statistics are approximate means over 6 repeats of the inference procedure. KL divergences: symmetrised Kullback-Leibler divergences ($\frac{1}{2} ( D_{KL}(P \| Q) + D_{KL}(Q \| P) )$, calculated with absolute discounting with constant $10^{-7}$) between a posterior $P$ from the given protocol and a posterior $Q$ using (i) `Approximate likelihood' and (ii) `Bare ABC $\epsilon = 0.1$' protocols.}}
\label{simstable}
}
\end{table}


\section{Experimental case study: transcription induction}

We now demonstrate the mean-first ABC approach by using it to infer parameters in a model of an experimentally measured biological system. For this case study, we use data on RNA transcript numbers from the well-known study on transcription rate by Golding \emph{et al.} \cite{golding2005real}. In this study, transcription was induced across a \reviewtwo{population of $\sim 100$ cells} at time $t=0$, and average RNA copy numbers were measured at several subsequent time points. This experiment was repeated \reviewtwo{$n = 3$} times, allowing a quantification of variance as well as mean copy number. As before, these observations of variance allow for more powerful inference, as the absolute values of parameters can be inferred as well as their difference.

In Ref. \cite{golding2005real}, the authors proposed a model to describe the behaviour of this system: after induction, cells have a constant probability $\lambda$ per unit time of producing a `burst' of RNA molecules, where burst size is geometrically distributed with average $\alpha$ (labelled $n$ in the original paper: we use $\alpha$ to avoid confusion with previous nomenclature). Cells grow and divide with cell cycle length $\tau$: at cell divisions, RNA molecules are binomially distributed between the two daughter cells. The population of cells is assumed to be unsynchronised, so at the start of the simulation, the time until the next cell division for any individual cell is a uniform random number on $[0, \tau]$. We also include a term $\nu$ representing a rate at which RNA molecules may degrade. The original paper considers overall rates of RNA production ($k_1$) and loss ($k_2$): the relationships between these variables are $k_1 = \lambda \alpha$ and $k_2 = \nu + \ln 2 / \tau$.

The mean copy number dynamics is straightforwardly obtained by considering the ODE

\begin{eqnarray}
\frac{d \mathbb{E}(x)}{dt} & = & k_1 - k_2 x \, ; \, x(t = 0) = 0; \\
\mathbb{E}(x) & = &  \frac{k_1}{k_2} \left(1 - e^{- k_2 t} \right).
\end{eqnarray}

We use stochastic simulation to estimate the variance associated with a given parameterisation. To recapitulate the experimental setup, each stochastic trajectory in our simulation consists of the mean copy number trajectory over a population of $100$ cells. We then perform an ensemble of $n$ of these stochastic trajectories to compute the statistics of the population-averaged copy number time series. 

\begin{figure*}
\includegraphics[width=18cm]{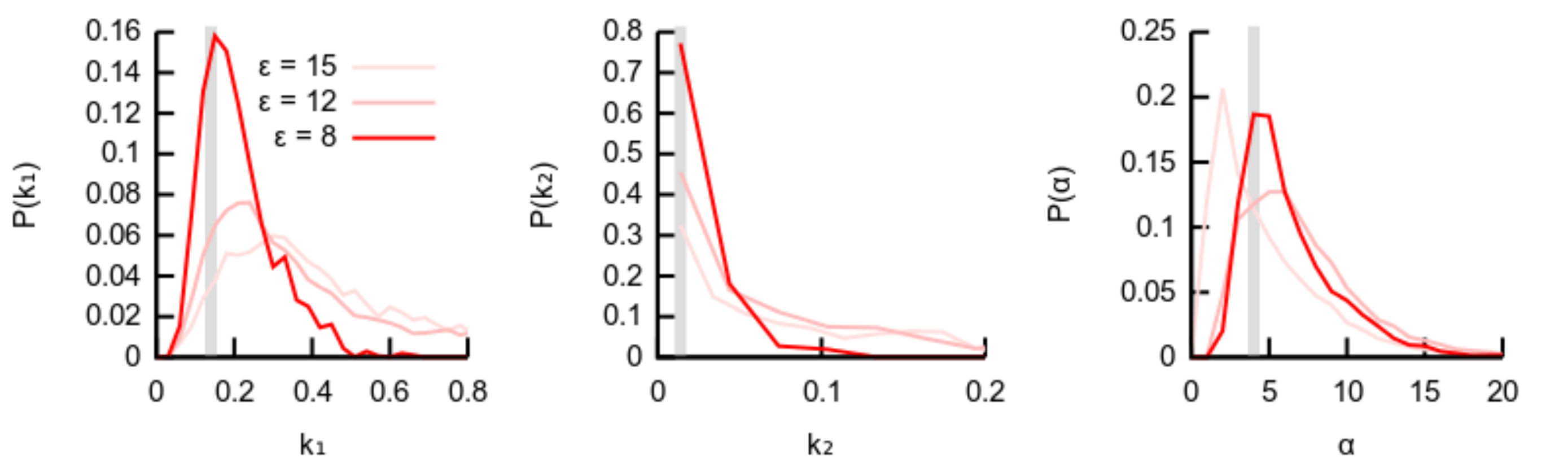}
\caption{\reviewtwo{\textbf{Mean-first ABC inferred posterior distributions for RNA induction experiments.} Inferred posterior distributions on (left) $k_1$, (centre) $k_2$ and (right) $\alpha$, the governing parameters for the model of transcription induction in Ref. \cite{golding2005real}. Thick vertical lines show the values assumed in the original study. The inferred posteriors show support for these values but uncertainty in these parameters, including pronounced skew in mean burst size $\alpha$, has now been quantified. \review{Posteriors resulting from different threshold values $\epsilon$ are shown to illustrate convergence.} }}
\label{figrnaposts}
\end{figure*}

Fig. \ref{figrnaposts} shows the posterior distributions that result from the mean-first ABC procedure with decreasing acceptance thresholds $\epsilon$. We \review{use uniform priors of} $k_1 \sim U(0, 0.5) \,\mbox{min}^{-1}$, $k_2 \sim U(0, 0.5) \,\mbox{min}^{-1}$, $\alpha \sim U(0, 100)$ (continuous, continous, and discrete, respectively). \review{As before, these priors are chosen to represent a range far wider than expected supported values, to ensure an equal prior probability of all values in this range, and to enforce non-negative restrictions on parameter ranges.} The posteriors on $k_1$, $k_2$, and $\alpha$ all exhibit density around the values chosen as representative in the original paper ($k_1 = 0.14 \,\mbox{min}^{-1}, k_2 = (\log{(2)} / 50)\,\mbox{min}^{-1}, \alpha = 4$) but uncertainties in these quantities have now been rigorously ascertained: in particular, the distribution of $\alpha$, the mean burst size, is rather skewed, suggesting that the experimental evidence supports the possibility of burst sizes at least twice as large as assumed in the original paper. \review{The posteriors increasingly converge as acceptance threshold $\epsilon$ is lowered.} Recording the number of trial parameterisations rejected by the mean-first ABC algorithm over 10 experimental repeats showed that in this case study, the implementation reduced the required number of stochastic simulations, and thus the computational load of the inference process, by around 25\%.


\begin{figure}
\includegraphics[width=9cm]{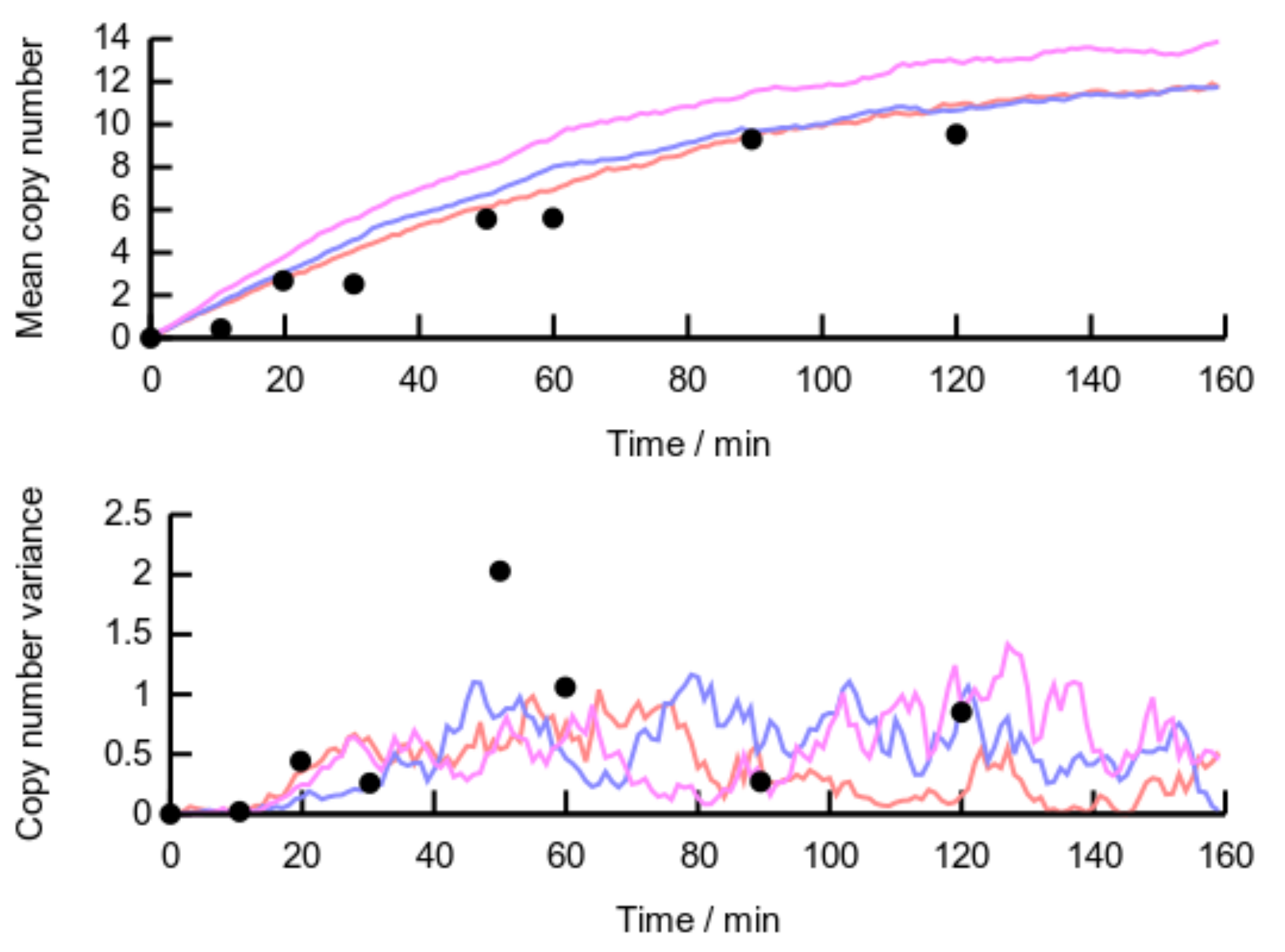}
\reviewtwo{\caption{\textbf{Example trajectories from mean-first ABC inference of RNA induction experiments.} (top) Mean and (bottom) variance of experimental \cite{golding2005real} and modelled RNA copy number after transcription induction. Points show the subset of measurement data used in the case study; lines show example trajectories with $\epsilon < 8$ resulting from trial parameterisations in the ABC inference simulations.}}
\label{figrnatrajs}
\end{figure}

\section{Discussion}
We have discussed methods for efficiently inferring the governing parameters of stochastic biological systems where means and variances are observed, including analytic forms for the likelihood associated with \review{sample} statistics, and ABC MCMC implementations for this class of problems. A method was introduced for making ABC MCMC more efficient by avoiding simulation of unimportant regions of parameter space, using an \review{upper bound of $\epsilon$ on the straightforwardly calculable mean trajectory discrepancy}. This method, when applied to synthetic and experimental data, reduced computational load by \reviewtwo{between 20\% and 50\%}. The mean-first protocol is completely generalisable to other computational frameworks than MCMC: it could equally well be employed in a simple approach sampling uniformly from the prior distribution (in which case it would likely yield a much more pronounced speedup due to the likely large volume of parameter space yielding unsuitable trajectories), or in other approaches including sequential Monte Carlo (SMC) which has also successfully been coupled to ABC \cite{sisson2007sequential}.


The algorithm we have presented here is conservative in its approach and only rejects parameterisations if the computed mean trajectory alone exceeds the threshold distance $\epsilon$. \review{In specific problems it may be possible to make this bound stricter. As an example, consider a case where it is known from the structure of a model that $\rho_v > \rho_m$ always. In this case, at least half of the total discrepancy $\rho_m + \rho_v$ will come from the contribution of the variance trajectory. An adjusted mean-first rejection protocol could be thus considered based on comparing computed mean trajectories to a refined threshold distance $\hat{\rho}_m < \epsilon/2$, as parameterisations that fail this initial comparison will likely exhibit $\rho_m + \rho_v > \epsilon$, given the contribution of the larger variance term. In such a case, yet more efficient rejection of unsuitable parameterisations may be achieved by forcing a stricter acceptance bound.}

In conclusion, we have presented analytic and computational methods to efficiently perform parametric inference in contexts where mean and variance measurements from a population of known or unknown size are available. The ability to quantitatively include variability in biological inference is not only important due to the physiological consequences of such variability but also allows more powerful inference of the mechanistic parameters of the system in question. We show that the computational approach we propose leads to substantial speedups in inference with both synthetic and experimental datasets and discuss how it may be coupled to other computational frameworks to suit the required context.

\review{
\section{Appendix}
We wish to show that if the discrepancy arising from the deterministic mean $\hat{\rho}_m$ exceeds $\epsilon$, it is likely that the combined discrepancies from the sample mean and sample variance $\rho_m + \rho_v$ also exceed $\epsilon$. \reviewtwo{We will assume that all measurements, and therefore the mean, are non-negative. Due to this non-negativity,} the relation $\hat{\rho}_m > \epsilon \implies \rho_m + \rho_v > \epsilon$ holds if $\hat{\rho}_m - \rho_m \leq \rho_v$. Consider the case in which the deterministic mean $\hat{\mu}$ differs by an amount $\delta$ from the \reviewtwo{expected value of the} sample mean $\mu$. The magnitude of $\delta$ is limited by the standard error on the mean: for a reasonably well-characterised mean measurement \reviewtwo{with low standard error}, we assume that $| \delta / \mu | \ll 1$. Expanding Eqns. \ref{detrhomean} and \ref{stochrhomean} gives}

\review{
\begin{eqnarray}
\hat{\rho}_m - \rho_m & = & (\log(\mu + \delta) - \log(m))^2 - (\log(\mu) - \log(m))^2 \\
& \simeq & ( \log(\mu) - \log(m) )^2 + 2 (\log(\mu) - \log(m)) \frac{\delta}{\mu} + \mathcal{O}( (\delta/\mu)^2 ) - ( \log(\mu) - \log(m) )^2 \\
& = & 2 (\log(\mu) - \log(m)) \frac{\delta}{\mu} + \mathcal{O}( (\delta/\mu)^2 ).
\end{eqnarray}
}

\review{If the discrepancy associated with mean measurements $(\log(\mu) - \log(m))$ is of a similar magnitude or less than the discrepancy from variance measurements $(\log(\sigma^2) - \log(v))$ then, neglecting higher-order terms, as $| \delta/\mu | \ll 1$, $2 (\log(\mu) - \log(m)) \delta / \mu \ll (\log(\sigma^2) - \log(v))^2$ and hence $\hat{\rho}_m - \rho_m < \rho_v$, the condition required for the validity of our threshold assumption. }



\bibliographystyle{unsrt}
\bibliography{refs1}

\begin{thebibliography}{10}

\bibitem{elowitz2002stochastic}
M.~B. Elowitz, A.~J. Levine, E.~D. Siggia, and P.~S. Swain.
\newblock {Stochastic gene expression in a single cell}.
\newblock {\em Science}, 297(5584):1183--1186, 2002.

\bibitem{kaern2005stochasticity}
M.~K{\ae}rn, T.~C. Elston, W.~J. Blake, and J.~J. Collins.
\newblock {Stochasticity in gene expression: from theories to phenotypes}.
\newblock {\em Nat. Rev. Genet.}, 6(6):451--464, 2005.

\bibitem{blake2003noise}
W.~J. Blake, M.~K{\ae}rn, C.~R. Cantor, and J.~J. Collins.
\newblock {Noise in eukaryotic gene expression}.
\newblock {\em Nature}, 422(6932):633--637, 2003.

\bibitem{raser2004control}
J.~M. Raser and E.~K. O'Shea.
\newblock {Control of stochasticity in eukaryotic gene expression}.
\newblock {\em Science}, 304(5678):1811--1814, 2004.

\bibitem{paulsson2005models}
J.~Paulsson.
\newblock {Models of stochastic gene expression}.
\newblock {\em Phys. Life Rev.}, 2(2):157--175, 2005.

\bibitem{rausenberger2008quantifying}
J.~Rausenberger and M.~Kollmann.
\newblock {Quantifying origins of cell-to-cell variations in gene expression}.
\newblock {\em Biophys. J.}, 95(10):4523--4528, 2008.

\bibitem{johnston2012mitochondrial}
I.~G. Johnston, B.~Gaal, R.~P. das Neves, T.~Enver, F.~J. Iborra, and N.~S.
  Jones.
\newblock Mitochondrial variability as a source of extrinsic cellular noise.
\newblock {\em PLoS computational biology}, 8(3):e1002416, 2012.

\bibitem{neves2010connecting}
R.~P. das Neves, N.~S. Jones, L.~Andreu, R.~Gupta, T.~Enver, and F.~J. Iborra.
\newblock {Connecting Variability in Global Transcription Rate to Mitochondrial
  Variability}.
\newblock {\em PLoS Biol.}, 8(12):451--464, 2010.

\bibitem{chang2008transcriptome}
H.~H. Chang, M.~Hemberg, M.~Barahona, D.~E. Ingber, and S.~Huang.
\newblock {Transcriptome-wide noise controls lineage choice in mammalian
  progenitor cells}.
\newblock {\em Nature}, 453(7194):544--547, 2008.

\bibitem{enver2009stem}
T.~Enver, M.~Pera, C.~Peterson, and P.~W. Andrews.
\newblock {Stem cell states, fates, and the rules of attraction}.
\newblock {\em Cell Stem Cell}, 4(5):387--397, 2009.

\bibitem{graf2008heterogeneity}
T.~Graf and M.~Stadtfeld.
\newblock {Heterogeneity of embryonic and adult stem cells}.
\newblock {\em Cell Stem Cell}, 3(5):480--483, 2008.

\bibitem{kussell2005bacterial}
E.~Kussell, R.~Kishony, N.~Q. Balaban, and S.~Leibler.
\newblock {Bacterial persistence: a model of survival in changing
  environments}.
\newblock {\em Genetics}, 169(4):1807, 2005.

\bibitem{brock2009non}
A.~Brock, H.~Chang, and S.~Huang.
\newblock {Non-genetic heterogeneity -- a mutation-independent driving force
  for the somatic evolution of tumours}.
\newblock {\em Nat. Rev. Genet.}, 10(5):336--342, 2009.

\bibitem{spencer2009non}
S.~L. Spencer, S.~Gaudet, J.~G. Albeck, J.~M. Burke, and P.~K. Sorger.
\newblock {Non-genetic origins of cell-to-cell variability in TRAIL-induced
  apoptosis}.
\newblock {\em Nature}, 459(7245):428--432, 2009.

\bibitem{wilkinson2012stochastic}
D.~J. Wilkinson.
\newblock {\em Stochastic modelling for systems biology}, volume~44.
\newblock CRC press, 2012.

\bibitem{gillespie1977exact}
D.T. Gillespie.
\newblock {Exact stochastic simulation of coupled chemical reactions}.
\newblock {\em J. Phys. Chem.}, 81(25):2340--2361, 1977.

\bibitem{toni2009approximate}
T.~Toni, D.~Welch, N.~Strelkowa, A.~Ipsen, and M.P.H. Stumpf.
\newblock Approximate bayesian computation scheme for parameter inference and
  model selection in dynamical systems.
\newblock {\em Journal of the Royal Society Interface}, 6(31):187--202, 2009.

\bibitem{beaumont2002approximate}
M.~A. Beaumont, W.~Zhang, and D.~J. Balding.
\newblock Approximate bayesian computation in population genetics.
\newblock {\em Genetics}, 162(4):2025--2035, 2002.

\bibitem{marin2012approximate}
J.-M. Marin, P.~Pudlo, C.~P. Robert, and R.~J. Ryder.
\newblock Approximate bayesian computational methods.
\newblock {\em Statistics and Computing}, 22(6):1167--1180, 2012.

\bibitem{sunnaker2013approximate}
Mikael Sunn{\aa}ker, Alberto~Giovanni Busetto, Elina Numminen, Jukka Corander,
  Matthieu Foll, and Christophe Dessimoz.
\newblock Approximate bayesian computation.
\newblock {\em PLoS computational biology}, 9(1):e1002803, 2013.

\bibitem{knight2000mathematical}
K.~Knight.
\newblock {\em Mathematical Statistics}.
\newblock Chapman \& Hall, 2000.

\bibitem{marjoram2003markov}
P.~Marjoram, J.~Molitor, V.~Plagnol, and S.~Tavar{\'e}.
\newblock Markov chain monte carlo without likelihoods.
\newblock {\em Proceedings of the National Academy of Sciences},
  100(26):15324--15328, 2003.

\bibitem{ding2008querying}
H.~Ding, G.~Trajcevski, P.~Scheuermann, X.~Wang, and E.~Keogh.
\newblock Querying and mining of time series data: experimental comparison of
  representations and distance measures.
\newblock {\em Proceedings of the VLDB Endowment}, 1(2):1542--1552, 2008.

\bibitem{bailey1964elements}
N.~T.~J. Bailey.
\newblock {\em The elements of stochastic processes with applications to the
  natural sciences}.
\newblock Wiley New York, 1964.

\bibitem{golding2005real}
I.~Golding, J.~Paulsson, S.~M. Zawilski, and E.~C. Cox.
\newblock Real-time kinetics of gene activity in individual bacteria.
\newblock {\em Cell}, 123(6):1025--1036, 2005.

\bibitem{sisson2007sequential}
S.~A. Sisson, Y.~Fan, and M.~M. Tanaka.
\newblock Sequential monte carlo without likelihoods.
\newblock {\em Proceedings of the National Academy of Sciences},
  104(6):1760--1765, 2007.

\end{thebibliography}

\end{document}